# Comment on
# "Wave Refraction in Negative-Index Media: Always Positive and Very Inhomogeneous"

JB Pendry[1] and D.R. Smith[2]

[1]*The Blackett Laboratory, Imperial College, London, SW7 2BU, UK.*
[2]*Department of Physics, UCSD, San Diego, CA 92093-0350, USA.*

In a recent Physical Review Letter [1] Valanju Walser and Valanju (VWV) called into question the basis of work on the so called negative index media (NIM). See for example [2,3]. The key point at issue is, 'what is the group velocity of a wave in NIM?'. The group velocity is central to the unusual properties claimed for these media. Everyone is in agreement with how the phase velocity refracts at the surface and that the phase velocity does show a negative index of refraction, as shown in figure 1(a).

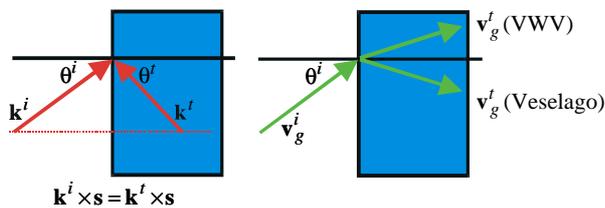

Figure 1. (a) Refraction of *wave fronts* at an interface between vacuum and a negative index medium (NIM). (b) Refraction of the *group velocity* at an interface between vacuum and a negative index medium (NIM).

Figure 1(b) shows the options for the group velocity: is the angle of refraction positive as VWV claim, or is it negative as Veselago claims? The question can speedily be resolved from the definition of group velocity,

$$\mathbf{v}_g = \nabla_\mathbf{k} \omega(\mathbf{k}) \qquad (1)$$

In all cases considered by VWV the NIM medium is isotropic so that $\omega$ does not depend on the direction of $\mathbf{k}$, only on the magnitude. Under these circumstances,

$$\mathbf{v}_g = \nabla_\mathbf{k} \omega(|\mathbf{k}|) = \mathbf{k}|\mathbf{k}|^{-1} d\omega(|\mathbf{k}|)/d|\mathbf{k}| \qquad (2)$$

The important point is that there are only two choices in an isotropic medium: $\mathbf{v}_g$ is either parallel or antiparallel to $\mathbf{k}$. Either way this result is not compatible with the conclusions of VWV. In fact we can deduce from Fig. 1 that $\mathbf{v}_g$ must be antiparallel to $\mathbf{k}$ because energy flow in the transmitted wave must always be away from the interface.

How did VWV come to a different conclusion? The problem is with the way they identify group velocity with an interference pattern. Two waves travelling in the same direction but at frequencies differing by $\delta\omega$ produce interference travelling with velocity $d\omega/dk = v_g$ which is identical in magnitude and direction to the group velocity; see figure 2(a). However in the calculation made by VWV the two waves on entering an NIM refract in slightly different directions, because of dispersion, to produce interference fronts that slide sideways in a crab like motion but with the true group velocity; see figure 2(b). VWV identified a *component* of $\mathbf{v}_g$ perpendicular to the interference fronts as the group velocity; Fig. 3 reconciles VWV's observations with the picture laid out by Veselago. Further details will appear in [4].

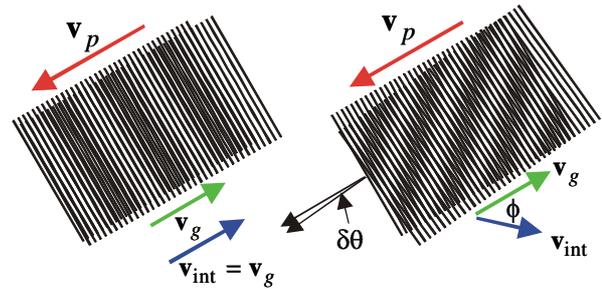

Figure 2. (a) In an isotropic medium two waves of different frequencies reveal the group velocity through their interference pattern *provided that* we choose the wave vectors of the two waves to be parallel. (b) Two waves of different frequencies and non parallel wave vectors result in an interference pattern moving with velocity $\mathbf{v}_{int}$ unrelated to the group velocity.

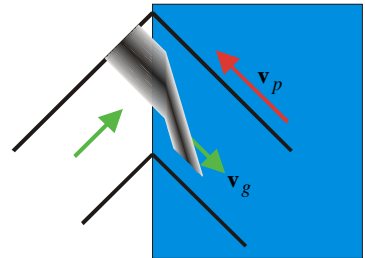

Figure 3. The phase fronts of a beam undergoing negative refraction.

To conclude: Veselago's result that both phase and group velocities undergo negative refraction at a vacuum/NIM interface is consistent with causality and with the well established properties of group velocity in isotropic media. VWV are correct when they calculate that interference fronts are positively refracted at a vacuum/NIM interface, but are wrong to interpret the normal to the interference front as the direction of the group velocity. The discrepancy can be resolved by noting that propagation of the front is crabwise, and antiparallel to the phase velocity, as required by Veselago.